\documentclass{article}
\pdfoutput=1
\usepackage{graphicx}
\usepackage{hyperref}
\usepackage{amsmath}
\usepackage{amsfonts}
\usepackage{amssymb}
\providecommand{\U}[1]{\protect\rule{.1in}{.1in}}
\newcommand{\Mvariable}[1]{}

\newcommand{\imag}[1]{\,i\,}
\begin{document}

\author{Graham G. Ross\thanks{g.ross@physics.ox.ac.uk\ } and Mario Serna
\thanks{serna@physics.ox.ac.uk\ }\\Rudolf Peierls Centre for Theoretical Physics,\\University of Oxford, 1 Keble Road, Oxford, OX1 3NP }
\title{Mass Determination of New States at Hadron Colliders}
\maketitle

\begin{abstract}
We propose an improved method for hadron-collider mass determination of new
states that decay to a massive, long-lived state like the LSP in the MSSM. We
focus on pair-produced new states which undergo three-body decay to a pair of
visible particles and the new invisible long-lived state. Our approach is to
construct a kinematic quantity which enforces all known physical constraints
on the system. The distribution of this quantity calculated for the observed
events has an endpoint that determines the mass of the new states. However we
find it much more efficient to determine the masses by fitting to the entire
distribution and not just the end point. We consider the application of the
method at the LHC for various models and demonstrate that the method can
determine the masses within about $6$ GeV using only $250$ events. This
implies the method is viable even for relatively rare processes at the LHC
such as neutralino pair production.
\end{abstract}

\section{Introduction}

At hadron colliders the determination of the masses of new particles
associated with missing momentum signals is very challenging due to the fact
that the kinematics of the event cannot be completely reconstructed. Hadron
colliders collide partons within each hadron, and each parton involved in the
collision carries an unknown fraction of the hadron's momentum. Therefore, the
center-of-mass (COM) energy and the frame of reference of the parton collision
are unknown for each event. The problem is further aggravated because one does
not expect any of the new short-lived particle states to travel far enough to
create tracks in the detector. In extensions of the Standard Model such as
supersymmetry (SUSY) or Universal Extra Dimensions (UED) there is often a
massive, stable, neutral particle that will leave the detector unnoticed,
leading to missing momentum associated with the production and
decay of the new particles required in such extensions.

For these reasons there has been much work developing techniques to determine
the mass of the new particles at hadron colliders such as the LHC. Significant
information comes from the endpoints of kinematic invariant distributions.
This is illustrated in the simple case that a short-lived state $Y$ undergoes
a three-body decay to a lepton pair plus the escaping neutral particle $N$
(the LSP in supersymmetry), $Y\rightarrow l^{+}+l^{-}+N. $ For this decay the
invariant mass $m_{ll}^{2}\equiv(k_{l^{-}}+k_{l^{+}})^{2}$ has a maximum value
equal to the mass difference $(M_{Y}-M_{N})^{2}$
\begin{equation}
\max m_{ll}=M_{Y}-M_{N}.\label{massdfference}%
\end{equation}
New states appear as bumps in the $m_{ll}$ distribution where one can read off
the mass difference from the upper edge of the bump \footnote{Loop corrections
play a role in shifting this endpoint slightly. For a detailed study see Ref
\cite{Drees:2006um}.  The shape is determined by the degree of interference with the slepton, see \cite{Phalen:2007te} for examples.  }. If $Y$ undergoes a two-body decay to an on-shell
intermediate state $Y\rightarrow X+l^{+}\rightarrow N+l^{+}+l^{-}$, then the
shape of the $m_{ll}$ distribution will be more like a right triangle with a
vertical drop, and the maximum $m_{ll}$ is given by $m^2_{ll}=(M_{Y}^{2}%
-M_{X}^{2})(M_{X}^{2}-M_{N}^{2})/M_{X}^{2}$. These techniques have been
extensively used to study SUSY in the context of the LHC (see Ref
\cite{Hinchliffe:1996iu} for a study of several models). Using events with
four leptons in the final states and missing energy, Ref \cite{Bisset:2005rn}
shows how such edges can form a Dalitz-like plot to determine information
about the mass spectra of new states. In short, such edges can accurately
determine relations between the masses of the unknown particles but not the
mass $M_{N}.$ The task of determining the complete mass spectra is
therefore dependent on determining $M_{N}$.

Much of the work in determining the $M_{N}$ in a hadron collider focuses on a
cascade of decays. The idea is to use events that contain many final states so
that one can find enough edges of invariant mass distributions to invert the
relationships and solve for the masses or SUSY model parameters. For example
Bachacou, Hinchliffe, Paige \cite{Bachacou:1999zb} use a sequence $\tilde{q}
\rightarrow q + \tilde{\chi}^{o}_{2} \rightarrow\tilde{l}^{-} + l^{+} + q
\rightarrow l^{-} + l^{+} + q + \tilde{\chi}^{o}_{1}$ involving four new
states in the event. One can form four invariant mass distributions from these
final states, and one has four unknown masses. This set of constraints
sometimes has multiple solutions. Fitting the shapes of the distributions can
lift this degeneracy as was shown in Miller, Osland, Raklev
(MOR)\cite{Miller:2005zp} and Lester \cite{Lester:2006yw}.

Is there a way to find $M_{N}$ if there are only three new states involved in
the event? Cheng, Gunion, Han, Marandella, McElrath (CGHMM)
\cite{Cheng:2007xv}
study pair-produced states, $Y,$ that decay via an on-shell intermediate state
$X$. An example scenario would be pair-produced $\tilde{\chi}_{2}^{o}s$ where
each branch decays via $\tilde{\chi}_{2}^{o}\rightarrow\tilde{l}^{+}%
+l^{-}\rightarrow l^{-}+l^{+}+\tilde{\chi}_{1}^{o}$ or its conjugate. Their
events consist of four leptons and missing energy. They analyze each event's
kinematics for compatibility with on-shell condition for the assumed topology.
To make their approach robust against background and finite resolution error,
they form distributions and use the shape to determine the unknown masses.

Finally, what can one determine from events which involve only two new states?
Cho Choi Kim and Park (CCKP)\cite{Cho:2007qv,Cho:2007dh} show how to use the
Cambridge transverse mass variable $M_{T2}$ of Lester and Summers
\cite{Lester:1999tx} to find $M_{Y}$ (which in their case was the gluino mass)
assuming a three-body decay to $\tilde{\chi}_{1}^{o}$ and $q$, $\bar{q} $.
Their example uses about 40000 events where gluinos are pair produced and
decay to four jets and missing energy. The $M_{T2}$ variable is a function
$\chi$ which is an assumed mass of $M_{N}$. One plots the maximum $M_{T2}%
(\chi)$ over the 40000 events as a function of $\chi$. A kink appears in the
function at the correct $M_{N}$ and $M_{Y}$ \footnote{For a recent study on
situations which lead to kinks using the transverse mass see ref
\cite{Barr:2007hy}}. Using this approach, CCKP find $M_{Y}$ and $M_{N}$ to
about $\pm2$ GeV for the case where $M_{+}/M_{-}\approx1.3$ where
\begin{equation}
M_{+}=M_{Y}+M_{N}\ \ \ \ M_{-}=M_{Y}-M_{N}.
\end{equation}

In this paper we will concentrate on the latter possibility involving the
production of only two new states. Our particular concern is to use the
available information as effectively as possible to reduce the number of
events needed to make an accurate determination of $M_{Y}$ and $M_{N}$. The
main new ingredient of the method proposed is that it does not rely solely on
the events close to the kinematic boundary but makes use of all the events.
Our method constrains the unobserved energy and momentum such that all the
kinematical constraints of the process are satisfied including the mass
difference, eq(\ref{massdfference}), which can be accurately measured from the
$ll$ spectrum. This increases the information that events far from the kinematic
boundary can provide about $M_{Y}$ and significantly reduces the number of
events needed to obtain a good measurement of the overall mass scale. Although
we develop the method for the case that $Y$ decays via a three-body decay to
an on-shell final states $Y\rightarrow N+l^{+}+l^{-},$ its
generalization to other processes is straightforward\footnote{We note that the
on-shell intermediate case studied by CGHMM is also improved by including the
relationship measured by the edge in the $ll$ distribution on each event's
analysis. The $Y$ decay channel with an on-shell intermediate state $X$ has an
edge in the $ll$ invariant mass distribution which provides a good
determination of the relationship $\max m_{ll}^{2}=(M_{Y}^{2}-M_{X}^{2}%
)(M_{X}^{2}-M_{N}^{2})/M_{X}^{2}$. This relationship forms a surface in
$M_{N}$,$M_{X}$,$M_{Y}$ space that only intersects the allowed points of
CGHMM's fig 3 near the actual masses.}.

In Section \ref{SecImprovedDistribution}, we introduce the $M_{2C}$
distribution whose endpoint gives $M_{Y}$, and whose distribution can be
fitted away from the endpoint to determine $M_{Y}$ and $M_{N}$ before one has
enough events to saturate the endpoint. Section \ref{SecEstimatedPerformance}
estimates the performance for a few SUSY models where we include approximate
detector resolution effects and where we expect backgrounds to be minimal.
Finally we conclude and discuss directions for further research. Appendix A
discusses the relationship between our distribution and the kink in
$M_{T2}(\chi)$ of CCKP and how this relationship can be used to find $M_{2C}$
in a computationally efficient manner. Appendix B provides details of our simulations.

\section{An improved distribution from which to determine $M_{Y}$}

\label{SecImprovedDistribution}

\begin{figure}[ptb]
\centerline{\includegraphics[width=3in]{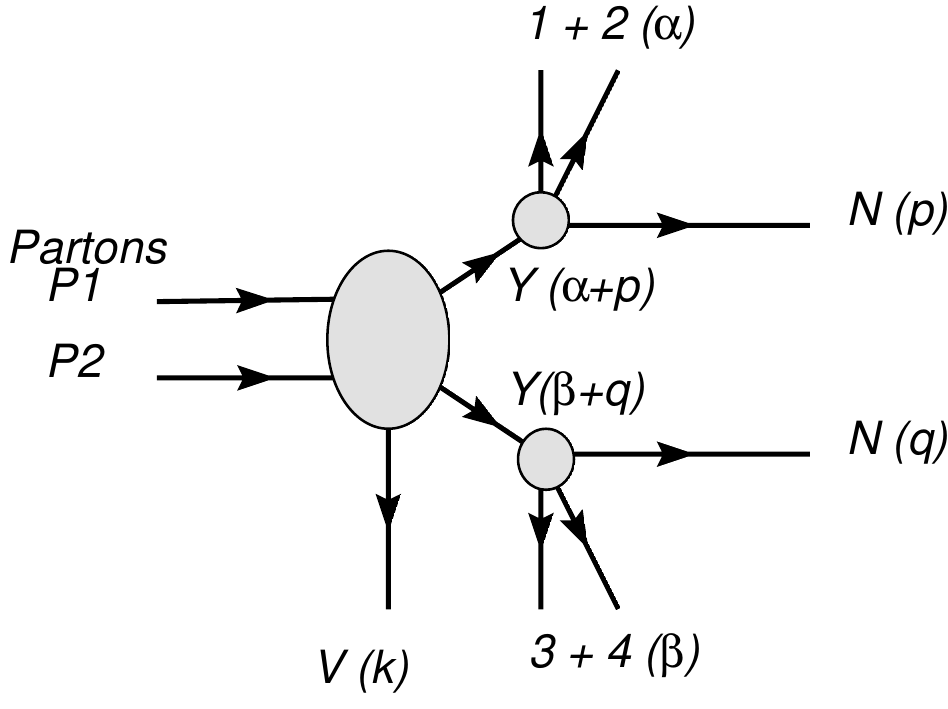}}\caption{We consider
events with the new state $Y$ is pair produced and in which each $Y$ decays
through a three-body decay to a massive state $N$ invisible to the detector
and visible particles $1$, $2$, $3$, and $4$.}%
\label{FigEventTopology}%
\end{figure}

We consider the event topology shown in fig \ref{FigEventTopology}. The new
state $Y$ is pair produced. Each branch undergoes a three-body decay to the
state $N$ with 4-momentum $p$ ($q$) and two visible particles $1+2$ ($3+4$)
with 4-momentum $\alpha$ ($\beta$). The invariant mass $m_{12}$ ($m_{34}$) of
the particles $1+2$ ($3+4$) will have an upper edge from which one can
well-determine $M_{-}$. Other visible particles not involved can be grouped
into $V$ with 4-momentum $k$. 
In the analysis presented here,
we assume $k=0$ and check that it remains valid for $k \lesssim 20$ GeV.

We adapt the concept from $M_{T2}$ of minimizing the transverse mass over the
unknown momenta to allow for the incorporation of all the available
information about the masses. To do this we form a new variable $M_{2C}$ which
we define as the minimum mass of the second to lightest new state in the event
$M_{Y}$ constrained to be compatible with the observed 4-momenta of $Y$'s
visible decay products with the observed missing transverse energy, with the
four-momenta of $Y$ and $N$ being on shell, and with the constraint that
$M_{-}=M_{Y}-M_{N}$ is given by the value determined by the end point of the
$m_{12}$ distribution.  The minimization is performed over the ten relevant unknown parameters which may be taken as the 4-momenta $p$ and $q$ of the states $N$, and the lab-frame collision energy $P_o$ and longitudinal momenta $P_z$. We neglect any contributions from unobserved initial state radiation
(ISR). Thus we have%
\begin{align}
M_{2C}^{2}  & =\min_{p,q,P_{o},P_{z}}(p+\alpha)^{2}\label{mymin}\\
& \mathrm{subject}\ \mathrm{to}\ \mathrm{the}\ 7\ \mathrm{constraints}%
\nonumber\\
& (p+\alpha)^{2}=(q+\beta)^{2},\\
& p^{2}=q^{2}\\
& (P_{o},0,0,P_{z})=p+q+\alpha+\beta+k\\
& \sqrt{(p+\alpha)^{2}}-\sqrt{(p)^{2}}=M_{-}.\label{EqDeltaMConstraint}
\end{align}
Although one can implement the minimization numerically or by using Lagrange
multipliers, we find the most computationally efficient approach is to modify
the $M_{T2}$ analytic solution from Lester and Barr \cite{Lester:2007fq}.
Details regarding implementing $M_{2C}$ and the relation of $M_{2C}$ to
$M_{T2}$ and the approach of CCKP are in Appendix A.

Errors in the determined masses propagated from the error in the mass
difference in the limit of $k=0$ are given by
\begin{equation}
\delta M_{Y} = \frac{\delta M_{-}}{2} \left(  1- \frac{M_{+}^{2}}{M_{-}^{2}}
\right)  \ \ \ \delta M_{N} = - \frac{\delta M_{-}}{2} \left(  1+ \frac
{M_{+}^{2}}{M_{-}^{2}} \right) \label{EqDeltaMmErrorEffects}%
\end{equation}
where $\delta M_{-}$ is the error in the determination of the mass difference
$M_{-}$. To isolate this source of error from those introduced by low
statistics, we assume we know the correct $M_{-}$, and one should consider the
error described in eq(\ref{EqDeltaMmErrorEffects}) as a separate uncertainty
from that reported in our initial performance estimates in the next section.

Because the true $p$, $q$, $P_{o}$, $P_{z}$ are in the domain over which we
are minimizing, $M_{2C}$ will always satisfy $M_{2C}\leq M_{Y}$. The equality
is reached for events with either $m_{12}$ or $m_{34}$ smaller than $M_{-},$
with $p_{z}/p_{o}=\alpha_{z}/\alpha_{o}$, and $q_{z}/q_{o}=\beta_{z}/\beta
_{o}$, and with the transverse components of $\alpha$ parallel to the
transverse components of $\beta$.

The events that approximately saturate the bound have the added benefit that
they are approximately reconstructed ($p$ and $q$ are known). If $Y$ is
produced near the end of a longer cascade decay, then this reconstruction
allows one to determine the masses of all the parent states in the event. The
reconstruction of several such events may also aid in spin correlation studies.

In order to determine the distribution of $M_{2C}$ for the process shown in
fig \ref{FigEventTopology}, we computed it for a set of events generated using
the theoretical cross section and assuming perfect detector resolution and no
background. Details of the simulation are in Appendix B. Figure
\ref{FigMYMinIdealExample} shows the resulting distribution for three cases:
$M_{Y}=200$ GeV, $M_{Y}=150$ GeV and $M_{Y}=100$ GeV each with $M_{-}=50$ GeV.
Each distribution was built from 30000 events. Note that the minimum $M_{Y}$
for an event is $M_{-}$. The endpoint in the three examples is clear, and one
is able to distinguish between different $M_{Y}$ for a given $M_{-}$.  The shape of the
distribution exhibits only modest model dependency as described in Appendix B.

One can also see that as $M_{+}/M_{-}$ becomes large, the $M_{Y}$
determination will be hindered by the small statistics available near the
endpoint or backgrounds. To alleviate this, one should instead fit to the
entire distribution. However it is clear that events away from the endpoint
also contain information about the masses. For this reason we propose to fit
the entire distribution of $M_{2C}$ and compare it to the `ideal' distribution
that corresponds to a given value of the masses. As we shall discuss this
allows the determination of $M_{Y}$ with a significant reduction in the number
of events needed. This is the most important new aspect of the method proposed here.

\begin{figure}[ptb]
\centerline{\includegraphics[width=6in]{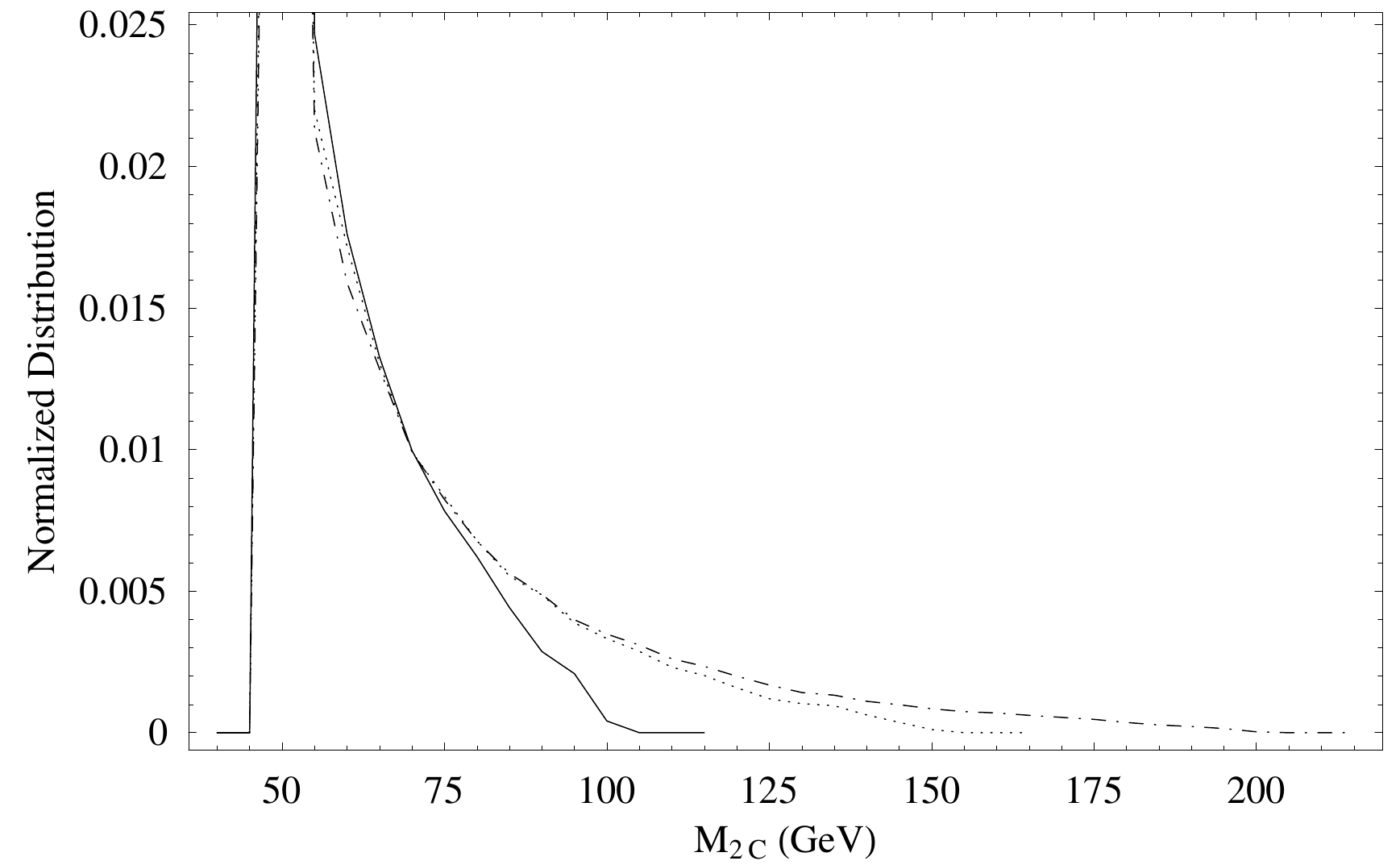}}\caption{The
distribution of 30000 events in 5 GeV bins with perfect resolution and no
background. The three curves represent $M_{Y}=200$ GeV (dot-dashed),
$M_{Y}=150$ GeV (dotted) and $M_{Y}=100$ GeV (solid) each with $M_{-}=50$ GeV.
Each distribution cuts off at the correct $M_{Y}$.}%
\label{FigMYMinIdealExample}%
\end{figure}

\section{Application of the method : SUSY model examples}

\label{SecEstimatedPerformance} To illustrate the power of the fit to the full
$M_{2C}$ distribution, we now turn to an initial estimate of one's ability to
measure $M_{Y}$ in a few specific supersymmetry scenarios. Our purpose here is
to show that fitting the $M_{2C}$ distribution can determine $M_{Y}$ and
$M_{N}$ with very few events. We include detector resolution effects but
neglect backgrounds 
but assume $k=0$ in the simulation. 
We calculate $M_{2C}$ for the case where the the analytic
$M_{T2}$ solution of Barr and Lester can be used to speed up the calculations
as described in Appendix A. Details on our calculations and simplifying
assumptions can be found in Appendix B. A more complete detailed study will
follow in a subsequent publication.

Although fitting the $M_{2C}$ distribution could equally well be applied to
the gluino mass studied in CCKP, we explore its applications to pair-produced
$\tilde{\chi}^{o}_{2}$. We select SUSY models where $\tilde{\chi}^{o}_{2}$
decays via a three-body decay to $l^{+} + l^{-} + \tilde{\chi}^{o}_{1}$. The
four momenta $\alpha=p_{l^{+}} + p_{l^{-}}$ for the leptons in the top branch,
and the four momenta $\beta=p_{l^{+}} + p_{l^{-}}$ for the leptons in the
bottom branch.


The production and decay cross section estimates in this section are
calculated using {MadGraph/MadEvent} \cite{Alwall:2007st}
and using SUSY mass spectra inputs from {SuSpect} \cite{Djouadi:2002ze}. The
distributions in this section still neglect background, but scale the $\alpha$
and $\beta$ four vectors by a scalar normally distributed about $1$ with the
width of
\begin{equation}
\frac{\delta\alpha_{0}}{\alpha_{0}} = \frac{0.1}{\sqrt{\alpha_{o}
(\mathrm{GeV})}} + \frac{0.003}{ \alpha_{o} (\mathrm{GeV})} + 0.007
\end{equation}
to simulate the typical LHC detector lepton energy resolution
\cite{AtlasTDR,CMSTDR}.   The missing transverse momentum is assumed to be whatever is missing to conserve the transverse momentum after the smearing of the leptons momenta.  We do not account for the greater uncertainty in missing momentum from hadrons or from muons which do not deposit all their energy in the calorimeter and whose energy resolution is therefore correlated to the missing momentum.  Including such effects requires a more detailed detector simulation and is beyond the scope of this Letter.    These finite resolution effects are simulated in the
determination of the ideal distribution and in the small sample of events that
is fit to the ideal distribution to determine $M_{Y}$ and $M_{N}$.  We do not expect expanded energy resolutions to greatly affect the results because the resolution effects are included in both the simulated events and in the creation of the ideal curves which are then fit to the low statistics events to estimate the mass.

We consider models where the three-body decay channel for $\tilde{\chi}%
_{2}^{o}$ will dominate. These models must satisfy $m_{\tilde{\chi}_{2}^{o}%
}-m_{\tilde{\chi}_{1}^{o}}<M_{Z}$ and must have all slepton masses greater
than the $m_{\tilde{\chi}_{2}^{o}}$. The models considered are shown in Table
\ref{TableModels}. The Min-Content model assumes that there are no other SUSY
particles accessible at the LHC other than $\tilde{\chi}_{2}^{o}$ and
$\tilde{\chi}_{1}^{o}$ and we place $m_{\tilde{\chi}_{1}^{o}}$ and
$m_{\tilde{\chi}_{2}^{o}}$ at the boundary of the PDG Live exclusion limit
\cite{PDBook2006}. SPS 6, P1, and $\gamma$ are models taken from references
\cite{Allanach:2002nj}, \cite{VandelliTesiPhD}, and \cite{DeRoeck:2005bw}
respectively. Each has the $\chi_{2}$ decay channel to leptons via a
three-body decay kinematically accessible. We will only show simulation results for the masses in model P1 and SPS 6 because they have the extreme values of $M_{+}/M_{-}$ with which the performance scales.   The Min-Content model and the $\gamma$ model are included to demonstrate the range of the masses and production cross sections that one might expect.

Bisset, Kersting, Li, Moortgat, Moretti, and Xie (BKLMMX) \cite{Bisset:2005rn}
have studied the 4 lepton + missing energy standard model background for the
LHC. They included contributions from jets misidentified as leptons and
estimated about $190$ background events at a ${\mathcal{L}}=300\ \mathrm{fb}%
^{-1}$ which is equivalent to $0.6$ fb. Their background study made no
reference to the invariant mass squared of the four leptons, so one only
expects a fraction of these to have both lepton pairs to have invariant masses
less than $M_{-}$. Their analysis shows the largest source of backgrounds will
most likely be other supersymmetric states decaying to four leptons. Again,
one expects only a fraction of these to have both lepton pairs with invariant
masses within the range of interest. The background study of BKLMMX is
consistent with a study geared towards a $500$ GeV $e^{+}$ $e^{-}$ linear
collider in ref \cite{Ghosh:1999ix} which predicts $0.4$ fb for the standard
model contribution to 4 leptons and missing energy.   The neutralino decay to $\tau$ leptons also provide a background because the $\tau$ decay to a light leptons $l=e,\mu$  ($\Gamma_{\tau \rightarrow l \bar{\nu}_l} / \Gamma \approx 0.34$) cannot be distinguished from prompt leptons. The neutrinos associated with these light leptons will be new sources of missing energy and will therefore be a background to our analysis.  The di-$\tau$ events will only form a background when both opposite sign same flavor $\tau$s decay to the same flavor of light lepton which one expects about 6\% of the time.

\begin{table}[ptb]%
\begin{tabular}
[c]{|c|c|c|c|c|}\hline
Model & Min Content (ref \cite{PDBook2006}) & SPS 6 (ref
\cite{Allanach:2002nj}) & P1 (ref \cite{VandelliTesiPhD}) & $\gamma$ ( ref
\cite{DeRoeck:2005bw})\\\hline
Definition &
\begin{tabular}
[c]{l}%
$\tilde{\chi}^{o}_{1}$ and $\tilde{\chi}^{o}_{2}$\\
are the only\\
LHC accessible\\
SUSY States\\
with smallest\\
allowed masses.
\end{tabular}
&
\begin{tabular}
[c]{l}%
Non Universal\\
Gaugino Masses\\
$m_{o}=150$ GeV\\
$m_{1/2} = 300$ GeV\\
$\tan\beta= 10$\\
$\mathrm{sign}(\mu) = +$\\
$A_{o}=0$\\
$M_{1}=480$ GeV\\
$M_{2}=M_{3}=300$ GeV
\end{tabular}
&
\begin{tabular}
[c]{l}%
mSUGRA\\
$m_{o}=350$ GeV\\
$m_{1/2} = 180$ GeV\\
$\tan\beta= 20$\\
$\mathrm{sign}(\mu) = +$\\
$A_{o}=0$%
\end{tabular}
&
\begin{tabular}
[c]{l}%
Non-Universal\\
Higgs Model\\
$m_{o} = 330$ GeV\\
$m_{1/2}=240$ GeV\\
$\tan\beta= 20$\\
$\mathrm{sign}(\mu) = +$\\
$A_{o}=0$\\
$H_{u}^{2} = -(242\,\mathrm{GeV})^{2}$\\
$H_{d}^{2} = +(373\,\mathrm{GeV})^{2}$\\
\end{tabular}
\\\hline
$m_{\tilde{\chi}^{o}_{1}}$ & $46$ GeV & $189$ GeV & $69$ GeV & $95$
GeV\\\hline
$m_{\tilde{\chi}^{o}_{2}}$ & $62.4$ GeV & $219$ GeV & $133$ GeV & $178$
GeV\\\hline
$M_{+}/M_{-}$ & $6.6$ & $13.6$ & $3.2$ & $3.3$\\\hline
\end{tabular}
\caption{Models with $\tilde{\chi}^{o}_{2}$ decaying via a three-body decay to
leptons.  We only show simulation results for the masses in model P1 and SPS 6 because they have the extreme values of $M_{+}/M_{-}$ with which the performance scales. }%
\label{TableModels}%
\end{table}

\begin{table}[ptb]%
\begin{tabular}
[c]{|c|c|c|c|}\hline
Model &
\begin{tabular}
[c]{l}%
$\sigma_{\tilde{\chi}^{o}_{2}\,\tilde{\chi}^{o}_{2}}$ Direct\\
$\sigma_{\tilde{\chi}^{o}_{2}\,\tilde{\chi}^{o}_{2}}$ Via $\tilde{g}$ or
$\tilde q$\\
\end{tabular}
&
\begin{tabular}
[c]{l}%
$\mathrm{BR}_{\tilde{\chi}^{o}_{2} \rightarrow l + \bar{l} + \tilde{\chi}%
^{o}_{1}}$\\
$\mathrm{BR}_{\tilde{\chi}^{o}_{2} \rightarrow q + \bar{q} + \tilde{\chi}%
^{o}_{1}}$%
\end{tabular}
&
\begin{tabular}
[c]{l}%
Events with\\
$4\,$ leptons $+ E_{T}$ missing\\
+ possible extra jets\\
${\mathcal{L}}=300\ \mathrm{fb}^{-1}$%
\end{tabular}
\\\hline
Min Content &
\begin{tabular}
[c]{l}%
$2130$ fb\\
N/A
\end{tabular}
&
\begin{tabular}
[c]{l}%
0.067\\
0.69
\end{tabular}
& 2893\\\hline
SPS 6 &
\begin{tabular}
[c]{l}%
$9.3$ fb\\
$626$ fb
\end{tabular}
&
\begin{tabular}
[c]{l}%
0.18\\
0.05
\end{tabular}
& 6366\\\hline
P1 &
\begin{tabular}
[c]{l}%
$35$ fb\\
$12343$ fb
\end{tabular}
&
\begin{tabular}
[c]{l}%
0.025\\
0.66
\end{tabular}
& 2310\\\hline
$\gamma$ &
\begin{tabular}
[c]{l}%
$17$ fb\\
$4141$ fb
\end{tabular}
&
\begin{tabular}
[c]{l}%
0.043\\
0.64
\end{tabular}
& 2347\\\hline
\end{tabular}
\caption{The approximate breakdown of signal events. }%
\label{TableEventCounts}%
\end{table}

\begin{figure}[ptb]
\centerline{\includegraphics[width=6in]{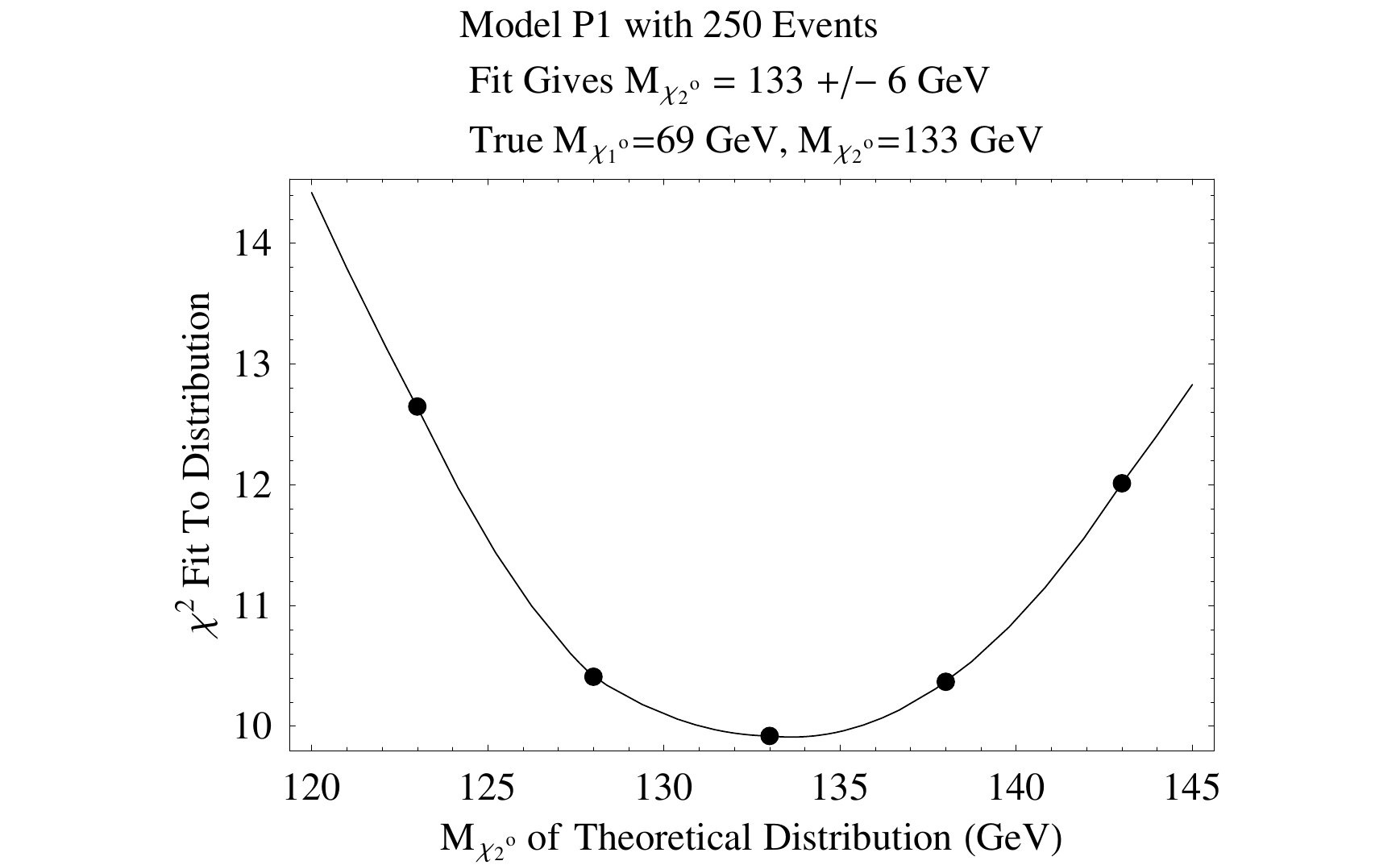}}\caption{$\chi^{2}$
fit of 250 events from model P1 of Ref \cite{VandelliTesiPhD} to the
theoretical distributions calculated for different $M_{\chi_{2}^{o}}$ values
but fixed $M_{\chi_{2}^{o}}-M_{\chi_{1}^{o}}$. \ The fit gives $M_{\chi
_{2}^{o}}=133\pm6$ GeV. }%
\label{FigP1ChiSqFitExample}%
\end{figure}

\begin{figure}[ptb]
\centerline{\includegraphics[width=6in]{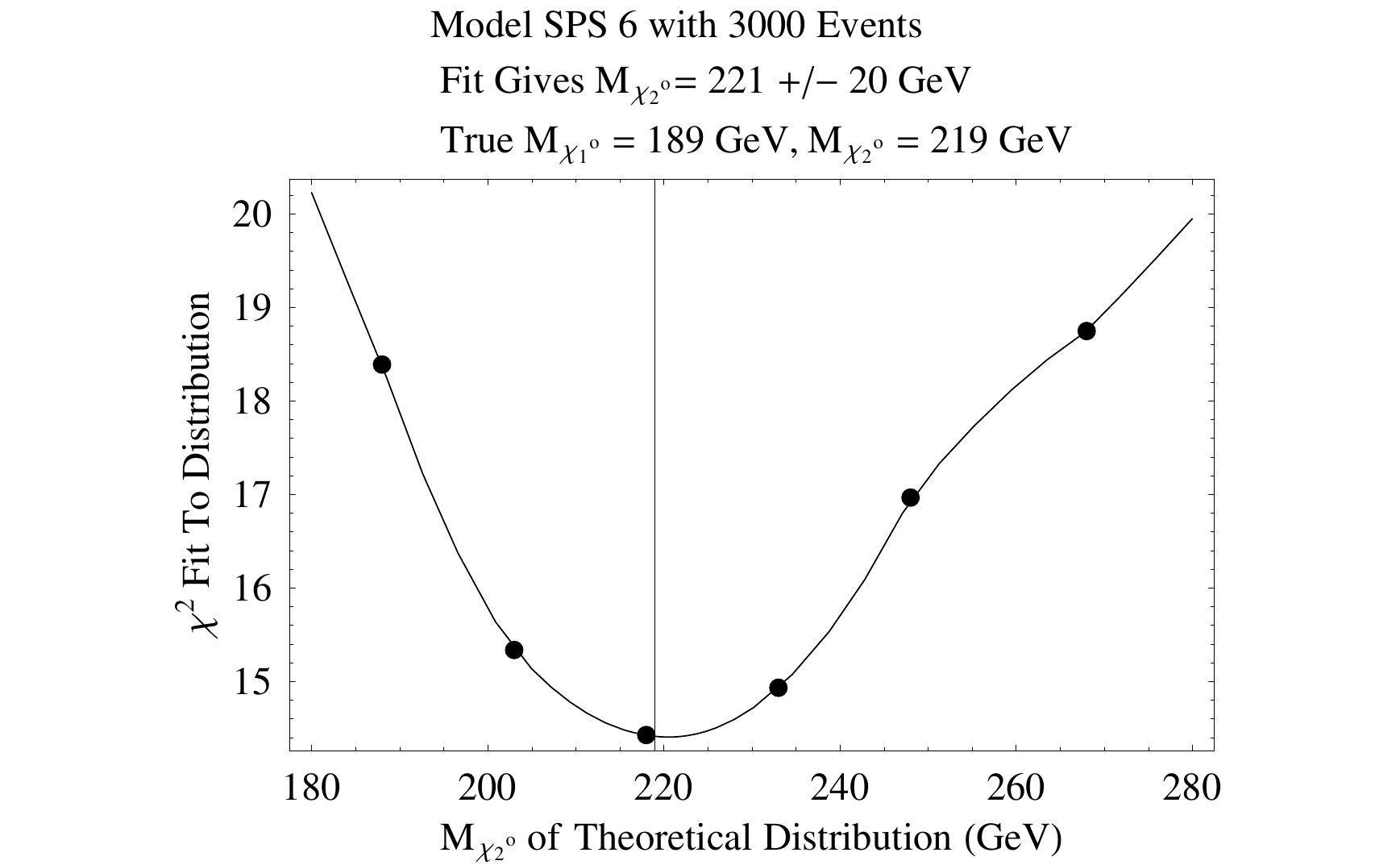}}\caption{$\chi^{2}$ fit
of 3000 events from model SPS 6 of Ref \cite{Allanach:2002nj} to the
theoretical distributions calculated for different $M_{\chi_{2}^{o}}$ values
but fixed $M_{\chi_{2}^{o}}-M_{\chi_{1}^{o}}$. The fit gives $M_{\chi_{2}^{o}%
}=221\pm20$ GeV. }%
\label{FigSPS6ChiSqFitExample}%
\end{figure}

Table \ref{TableEventCounts} breaks down the LHC production cross section for
pair producing two $\tilde{\chi}^{o}_{2}$ in each of these models. In the
branching ratio to leptons, we only consider $e$ and $\mu$ states as the
$\tau$ will decay into a jet and a neutrino introducing more missing energy.
Direct pair production of $\tilde{\chi}^{o}_{2}$ has a rather modest cross
section, but production via a gluino or squark has a considerably larger cross
section but will be accompanied by additional QCD jets. One does expect to be
able to distinguish QCD jets from $\tau$ jets \cite{2005NuPhS.144..341T}.

We now estimate how well one may be able to measure $m_{\tilde{\chi}_{1}^{o}}
$ and $m_{\tilde{\chi}_{2}^{o}}$ in these models. Figures
\ref{FigP1ChiSqFitExample} and \ref{FigSPS6ChiSqFitExample} show a $\chi^{2}$
fit\footnote{See Appendix B for details of how $\chi^{2}$ is calculated.} of
the $M_{2C}$ distribution from the observed small set of events to `ideal'
theoretical $M_{2C}$ distributions parameterized by $m_{\tilde{\chi}_{2}^{o}}%
$. The `ideal' theoretical distributions are calculated for the observed value
of $M_{-}$ using different choices for $m_{\tilde{\chi}_{2}^{o}}$. A
second-order interpolation is then fit to these points to estimate the value
for $m_{\tilde{\chi}_{2}^{o}}$. The $1\,\sigma$ uncertainty for $m_{\tilde
{\chi}_{2}^{o}}$ is taken to be the points where the $\chi^{2}$ increases from
its minimum by one.

The difficulty of the mass determination from the distribution grows with the
ratio $M_{+}/M_{-}.$ Figures \ref{FigP1ChiSqFitExample} and
\ref{FigSPS6ChiSqFitExample} show the two extremes among the cases we
consider. For the model P1 $M_{+}/M_{-}=3.2$, and for model $\gamma$
$M_{+}/M_{-}=3.3$. Therefore these two models can have the mass of
$m_{\tilde{\chi}_{2}^{o}}$ and $m_{\tilde{\chi}_{1}^{o}}$ determined with
approximately equal accuracy with equal number of signal events. Figure
\ref{FigP1ChiSqFitExample} shows that one may be able to achieve $\pm6$ GeV
resolution after about $30\ \mathrm{fb}^{-1}$. Model SPS 6 shown in fig
\ref{FigSPS6ChiSqFitExample} represents a much harder case because
$M_{+}/M_{-}=13.6$. In this scenario one can only achieve $\pm20$ GeV
resolution with 3000 events corresponding to approximately $150\,\mathrm{fb}%
^{-1}$. In addition to these uncertainties, one needs to also consider the
error propagated from $\delta M_{-}$ in eq(\ref{EqDeltaMmErrorEffects}).


\section{Summary and Conclusions}

\label{SecConclusions}

We have proposed a method to extract the masses of new pair-produced states
based on a kinematic variable, $M_{2C}$, which incorporates all the known
kinematic constraints on the observed process and whose endpoint determines
the new particle masses. However the method does not rely solely on the
endpoint but uses the full data set, comparing the observed distribution for
$M_{2C}$ with the ideal distribution that corresponds to a given mass. As a
result the number of events needed to determine the masses is very
significantly reduced so that the method may be employed at the LHC event for
processes with electroweak production cross sections.

We have performed an initial feasibility study of the method for several
supersymmetric models. This includes the effect of detector resolution but not
backgrounds, cuts and combinatoric complications
but was modeled with an assumption that $k=0$. We demonstrated that for
some of the models studied we are able to determine the masses to within 6 GeV
from only 250 events. This efficiency is encouraging although a study
including more of the real-world complications is needed to augment this
initial study.

The method we advocate here can be readily extended to other processes. By
incorporating all the known kinematical constraints, the information away from
kinematical end-points can, with some mild process dependent information, be
used to reduce the number of events needed to get mass measurements. We shall
illustrate this for other cases elsewhere[in preparation].

\section*{Acknowledgements}

We would like to thank Alan Barr for many stimulating insights and for
reviewing the first drafts of the paper. We also want to thank James Gray,
Chris Lester, Tilman Plehn, John March Russell, and Laura Serna for helpful
conversations. We owe thanks to Fabio Maltoni and Tim Stelzer for providing us
online access to MadGraph and MadEvent tools. M.S. acknowledges support from
the United States Air Force Institute of Technology. The views expressed in
this letter are those of the authors and do not reflect the official policy or
position of the United States Air Force, Department of Defense, or the US Government.

\section*{Appendix A : Using $M_{T2}$ to Find $M_{2C}$}

\label{SecAppendixRelateToMT2} The variable $M_{T2}$, which was introduced in
by Lester and Summers \cite{Lester:1999tx}, is equivalent to
\begin{align}
M_{T2}^{2}(\chi)  & =\min_{p,q,P_{o},P_{z}} (p+\alpha)^{2}\label{mt2}\\
& \mathrm{subject}\ \mathrm{to}\ \mathrm{the}\ 7\ \mathrm{constraints}%
\nonumber\\
& (p+\alpha)^{2}=(q+\beta)^{2},\\
& p^{2}=q^{2}\\
& (P_{o},0,0,P_{z})=p+q+\alpha+\beta+k\\
& p^{2}=\chi^{2}.\label{EqChiConstraint}%
\end{align}
As is suggested in the simplified example of \cite{Gripaios:2007is}, the
minimization over $P_{o}$ and $P_{z}$ is equivalent to assuming $p$ and
$\alpha$ have equal rapidity and $q$ and $\beta$ have equal rapidity.
Implementing this eq(\ref{mt2}) reduces to the traditional definition of the
Cambridge transverse mass.

\begin{figure}[ptb]
\centerline{\includegraphics[width=4in]{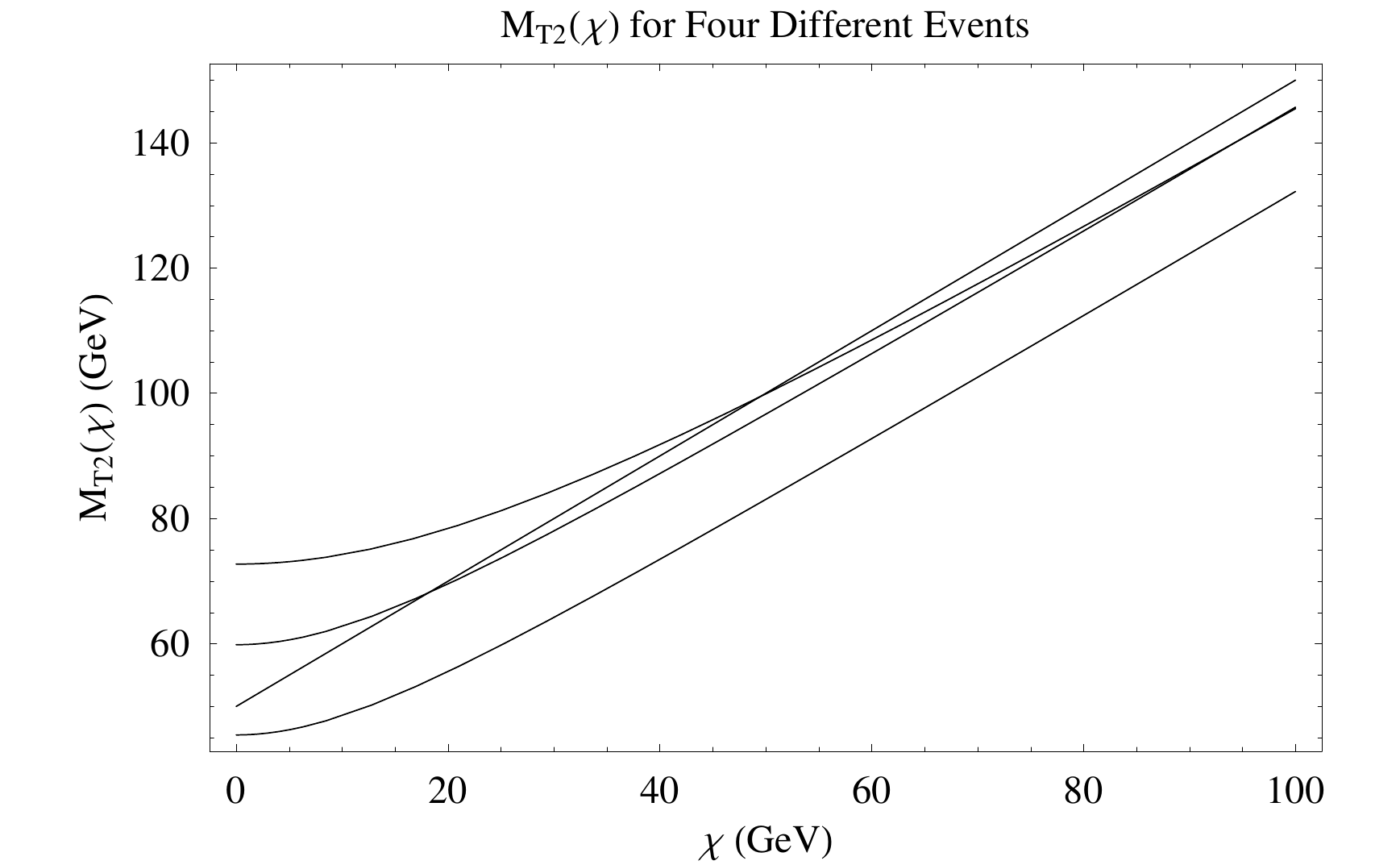}}\caption{ The
$M_{T2}(\chi)$ curves for four events with $M_{N}=50$ GeV and $M_{Y}=100$ GeV.
Only the events whose curves starts off at $M_{T2}(0) > M_{-}$ intersect the
straight line given by $M_{T2}(\chi) - \chi= M_{-}$. The $M_{T2}$ at the
intersection is $M_{2C}$ for that event. }%
\label{FigMT2ComparisonPlot}%
\end{figure}

By comparing $M_{T2}(\chi)$ as defined above to $M_{2C}$ defined in
eq(\ref{mymin}), one can see that they are very similar with the exception
that the constraint eq(\ref{EqDeltaMConstraint}) is replaced by the constraint
eq(\ref{EqChiConstraint}). $M_{2C}$ can be found by scanning $M_{T2}(\chi)$
for the $\chi$ value that such that the constraint in
eq(\ref{EqDeltaMConstraint}) is also satisfied.

One can see the $M_{2C}$ and $M_{T2}$ relationship visually. Each event
provides a curve $M_{T2}(\chi)$; fig \ref{FigMT2ComparisonPlot} shows curves
for four events with $M_{N}=50$ GeV and $M_{Y}=100$ GeV. For all events
$M_{T2}(\chi)$ is a continuous and monotonically increasing function of $\chi$. As CCKP point
out, at large $\chi$ 
 and at $k=0$ the \textit{maximum }$M_{T2}(\chi)$ approaches
$\chi+M_{-}$ so one knows the slope of $M_{T2}(\chi)$ for all events will be
everywhere less than or equal to one. Furthermore if $M_{T2}( \chi=0)>M_{-},$
as is true for two of the four events depicted in fig
\ref{FigMT2ComparisonPlot}, then, barring an asymptote, there is a solution to $M_{T2}(\chi
)=\chi+M_{-}$. At this point $M_{T2}(\chi)=\min M_{Y} |_{\mathrm{Constraints}}
\equiv M_{2C}$. Equivalently
\begin{align}
M_{2C}  & =M_{T2}\ \ \mathrm{at}\ \chi\ \mathrm{where}\ \ M_{T2}(\chi
)=\chi+M_{-}\ \ \ \mathrm{if}\ \ M_{T2}(\chi=0)>M_{-}\\
& =M_{-}\ \ \ \ \mathrm{otherwise}.
\end{align}
At $k=0$ the maximum $\chi$ of such an intersection occurs for $\chi=M_{N}$ which is
why the endpoint of $M_{2C}$ occurs at the correct $M_{Y}$ and why this
corresponds to the kink of CCKP. Because Barr and Lester have an analytic
solution to $M_{T2}$ in ref \cite{Lester:2007fq} in the case $k=0$, this is
computationally very efficient as a definition.

\section*{Appendix B: Numerical Simulation Details}

\subsection*{Numerical simulation of \textquotedblleft ideal\textquotedblright%
\ events}

\label{SecAppendixSimulationDetails}

In order to determine the distribution of $M_{2C}$ for the processes shown in
fig \ref{FigEventTopology}, it is necessary to generate a large sample of
\textquotedblleft ideal\textquotedblright\ events corresponding to the
physical process shown in the figure. For simplicity in the numerical
simulations included in this note we always assume $k=0$ and decay via an
off-shell Z-boson as this is what could be calculated quickly and captures the
essential elements to provide an initial estimate of our approach's utility.

Even under these assumptions, one might expect that the shape of distribution
depends sensitively on the parton distribution and many aspects of the
differential cross section and differential decay rates. Surprisingly this is
not the case; the shape of the distribution depends sensitively only on two properties:

(i) the shape of the $m_{12}$ (or equivalently $m_{34}$) distributions. In the
examples studied here for illustration we calculate the $m_{12}$ distribution
assuming it is generated by a particular supersymmetric extension of the Standard Model, but in practice one should use the measured distribution which is accessible to accurate determination.  The particular shape of $m_{12}$ does not greatly affect the ability to determine the mass of $N$ and $Y$ so long as one can still find the endpoint to determine $M_Y-M_N$ and use the observed $m_{ll}$ distribution to model the shape of the $M_{2C}$ distribution.

(ii) the angular dependence of the $N$'s momenta in the rest frame of $Y$. In
the preliminary analysis presented here we assume that in the rest frame of
$\tilde{\chi}_{2}^{o}$, $\tilde{\chi}_{1}^{o}$'s momentum is distributed
uniformly over the $4\pi$ steradian directions. While this assumption is not
universally true it applies in many cases and hence is a good starting point
for analyzing the efficacy of the method.

Under what conditions is the uniform distribution true? Note that the
$\tilde{\chi}_{2}^{o}$'s spin is the only property of $\tilde{\chi}_{2}^{o}$
that can break the rotational symmetry of the decay products. For $\tilde
{\chi}_{2}^{o}$'s spin to affect the angular distribution there must be a
correlation of the spin with the momentum which requires a parity violating
coupling. Consider first the Z contribution. Since one is integrating over the
lepton momenta, the parity violating term in the cross section coming from the
lepton-Z vertex vanishes and a non-zero correlation requires that the parity
violating coupling be associated with the neutralino vertex. The Z-boson
neutralino vertex vanishes as the Z interaction is proportional to
$\overline{\tilde{\chi}_{2}^{o}}\gamma^{5}\gamma^{\mu}\tilde{\chi}_{1}%
^{o}Z_{\mu}$ or $\overline{\tilde{\chi}_{2}^{o}}\gamma^{\mu}\tilde{\chi}%
_{1}^{o}Z_{\mu}$ depending on the relative sign of $m_{\tilde{\chi}_{2}^{o}}$
and $m_{\tilde{\chi}_{1}^{o}}$ eigenvalues. However if the decay has a
significant contribution from an intermediate slepton there are parity
violating couplings and there will be spin correlations. In this case there
will be angular correlations but it is straightforward to modify the method to
take account of correlations. We hope to study this in another
publication\footnote{Studying and exploiting the neutralino spin correlations
is discussed further in Refs
\cite{MoortgatPick:1999di,MoortgatPick:2000db,Choi:2005gt}.}.

Even in the case that the slepton contribution is significant the correlations
may still be absent. Because we are worried about a distribution, the spin
correlation is only of concern to our assumption if a mechanism aligns the
spin's of the $\tilde{\chi}_{2}^{o}$s in the two branches. Table
\ref{TableEventCounts} shows that most of the $\tilde{\chi}_{2}^{o}$ one
expects follow from decay chains involving a squark, which being a scalar
should make uncorrelated the spin of the $\tilde{\chi}_{2}^{o} $ in the two
branches. One would then average over the spin states of $\tilde{\chi}_{2}%
^{o}$ and recover the uniform angular distribution of $\tilde{\chi}_{1}^{o}$'s
momentum in $\tilde{\chi}_{2}^{o}$'s rest frame.

\begin{figure}[ptb]
\centerline{
\includegraphics[width=3.2in]{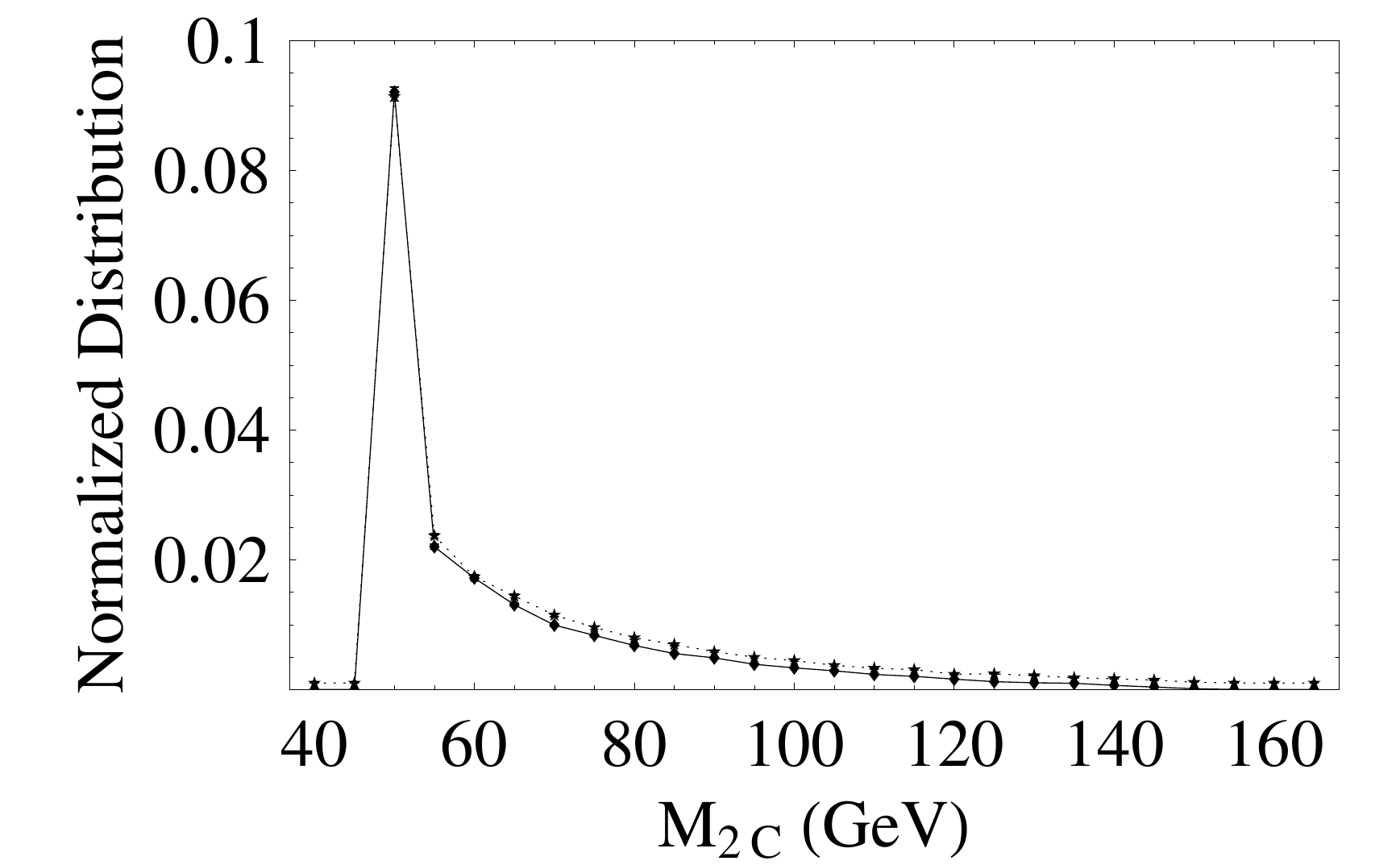}
\includegraphics[width=3.2in]{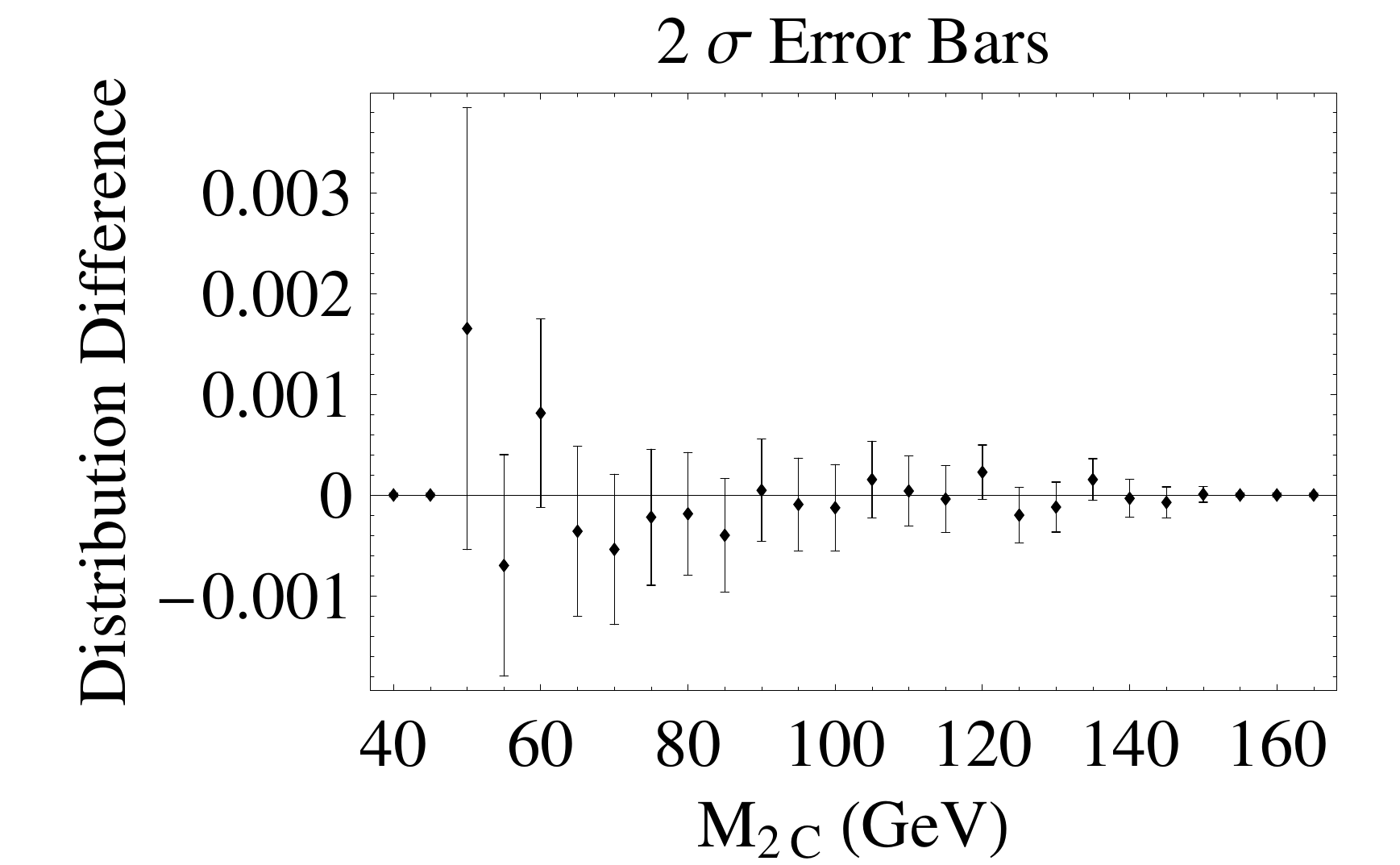}}\caption{Demonstrates
the distribution is independent of the COM energy, angle with which the pair
is produced with respect to the beam axis, and the frame of reference.}%
\label{FigShapeIndenpendence}%
\end{figure}

Once one has fixed the dependencies (i) and (ii) above, the shape of the
distribution is essentially independent of the remaining parameters. To
illustrate this result we show in fig \ref{FigShapeIndenpendence} two cases:

(1) The case that the collision energy and frame of reference and angle of the
produced $Y$ with respect to the beam axis are distributed according to the
calculated cross section for the process considered in Section
\ref{SecEstimatedPerformance} in which $\tilde{\chi}_{2}^{o}$ decays via $Z$
exchange to the three-body state $l^{+}+l^{-}+\tilde{\chi}_{1}^{o},$
convoluted with realistic parton distribution functions.

(2)The case that the angle of the produced $Y$ with respect to the beam axis
is arbitrarily fixed at $\theta=0.2$ radians, the azimuthal angle $\phi$ fixed
at $0$ radians, and the total 4-momentum of the colliding particles
arbitrarily set to $P=(500,0,0,0)$ GeV.

The left plot of fig \ref{FigShapeIndenpendence} shows the two distributions
intentionally shifted by 0.001 to allow one to barely distinguish the two
curves. On the right side of fig \ref{FigShapeIndenpendence} we show the
difference of the two distributions with the 2 $\sigma$ error bars within
which one expects 95\% of the bins to overlap $0$ if the distributions are
identical. In addition 
to tests with $k=0$, we also tested that $k \lesssim 20$ GeV does not change the shape of
the distribution 
to within our numerical uncertainties for 
any of our results. In a test case  
where we constructed events with $M_Y=150$ GeV and $M_N=100$ GeV, with $\sqrt{k^2}$ uniformly distributed with between $2$ of $20$ GeV, with $|\vec{k}/k_0| = 0.98$, and with uniform angular distribution, we
found the $M_{2C}$ distribution agreed with the distribution shown in 
figure \ref{FigShapeIndenpendence} within the expected error bars after 10000 events. 
Scaling this down to the masses studied in $P1$ we trust these results remain unaffected for $k \lesssim 20$ GeV. 
Introduction of
cuts on jets and missing traverse energy will probably introduce some
dependence on the COM energy of the collision that is absent in this ideal case.

Given this structure detailed in (i) and (ii) above we calculate the
\textquotedblleft ideal\textquotedblright\ distributions for $M_{2C}$ assuming
that $k=0$ and that in the rest frame of $Y$ there is an equal likelihood of
$N$ going in any of the $4\pi$ steradian directions. The observable invariant
$\alpha^{2}$ is determined according to the differential decay probability of
$\chi_{2}^{o}$ to $e^{+}$ $e^{-}$ and $\chi_{1}^{o}$ through a Z-boson
mediated three-body decay. Analytic expressions for cross sections were
obtained from the Mathematica output options in {CompHEP} \cite{Boos:2004kh}

Inclusion of backgrounds will change the shape. Backgrounds that one can anticipate or measure, like di-$\tau$s or leptons from other neutralino decays observed with different edges can be modeled and included in the ideal shapes used to perform the mass parameter estimation.  A more complete study is beyond the scope of this letter and will follow in a subsequent publication.

\subsection*{Least squares fit}

In order to determine $M_{Y}$ it is necessary to quantify the comparison
between the $N$ observed events and the \textquotedblleft
ideal\textquotedblright\ events. To do this we define a $\chi^{2}$
distribution by computing the number of events, $C_{j},$ in a given range,
$j,$ (bin $j$) of $M_{2C}.$ Assuming a Poisson distribution, we assign an
uncertainty, $\sigma_{j}$, to each bin $j$ given by
\begin{equation}
\sigma_{j}^{2}=\frac{1}{2}\left(  N\,f({M_{2C}}_{j},M_{Y}) + C_{j}\right)  .
\end{equation}
Here the normalized distribution of ideal events is $f(M_{2C},M_{Y})$, and the
second term has been added to ensure that the contribution of bins with very
few events, where Poisson statistics does not apply\footnote{By this we mean
that $N\,f({M_{2C}}_{j},M_{Y})$ has a large percent error when used as a
predictor of the number of counts $C_{j}$ when $N\,f({M_{2C}}_{j},M_{Y})$ is
less than about 5.}, have a reasonable weighting. Then $\chi^{2}$ is given by
\begin{equation}
\chi^{2}(M_{Y})=\sum_{\mathrm{bin}\ j}  \left(  \frac
{C_{j}-N\,f({M_{2C}}_{j},M_{Y})}{\sigma_{j}}\right)  ^{2} .
\end{equation}
The minimum $\chi^{2}(M_{Y})$ is our estimate of $M_{Y}$. The amount $M_{Y}$
changes for an increase of $\chi^{2}$ by one gives our $1\,\sigma$
uncertainty, $\delta M_{Y}$, for $M_{Y}$ \cite{Bevington}. As justification
for this we calculate ten different seed random numbers to generate ten
distinct groups of 250 events. We check that the $M_{Y}$ estimates for the ten
sets are distributed with about 2/3 within $\delta M_{Y}$ of the true $M_{Y}$
as one would expect for $1\,\sigma$ error bars. One might worry that with our
definition of $\chi^{2}$, the value of $\chi^{2}$ per degree of freedom is
less than one. However this is an artifact of the fact that the bins with very
few or zero events are not adequately described by Poisson statistics and if
we remove them we do get a reasonable $\chi^{2}$ per degree of freedom. The
determination of $M_{Y}$ using this reduced set gives similar results.


\providecommand{\href}[2]{#2}\begingroup\raggedright

\endgroup

\end{document}